\documentclass{pas}
\usepackage{multirow,natbib,aas_macros}
\usepackage{gensymb}
\usepackage{amsmath}
\usepackage{graphicx}
\usepackage{float}
\usepackage[T1]{fontenc}

\begin{document}

\newcommand{\chandra}{\textit{Chandra}}
\newcommand{\acis}{\textit{Chandra}/ACIS}
\newcommand{\hrc}{\textit{Chandra}/HRC}

\lefttitle{Publications of the Astronomical Society of Australia}
\righttitle{Cambridge Author}

\jnlPage{1}{4}
\jnlDoiYr{2021}
\doival{10.1017/pasa.xxxx.xx}

\articletitt{Research Paper}

\title{A new likely pulsar binary in 47~Tucanae from continuum searches}

\author{\sn{Thomas J.} \gn{Maccarone}$^{1}$, \sn{Alessandro} \gn{Paduano}$^{2}$, \sn{Arash} \gn{Bahramian}$^2$, \sn{Liliana} \gn{Rivera Sandoval}$^{3}$, \sn{Jay} \gn{Strader}$^4$, \sn{James C.A.} \gn{Miller-Jones}$^2$, \sn{Angiraben D.} \gn{Mahida}$^2$, \sn{Craig O.} \gn{Heinke}$^5$, \sn{Laura} \gn{Chomiuk}$^4$ }
\affil{$^1${Texas Tech University, USA }$^2$ {International Centre for Radio Astronomy Research -- Curtin University}, $^3$ {University of Texas-Rio Grande Valley, USA}, $^4$ {Michigan State University, USA}, $^5$ {University of Alberta, Edmonton, Canada} }

\corresp{T.J. Maccarone, Email: thomas.maccarone@ttu.edu}
\history{(Received xx xx xxxx; revised xx xx xxxx; accepted xx xx xxxx)}

\begin{abstract}
We present evidence that the X-ray source 47~Tuc~W41, long considered to be an X-ray source powered by coronal activity, is actually a redback pulsar binary. The source continually shows $L_X = 3 \times 10^{31}$ erg s$^{-1}$ (0.5--10 keV), which is well in excess of the coronal saturation limit for active binaries in quiescence, but is consistent with the intrabinary shock observed in redback pulsars, as is its photon index of $\Gamma = 1.4\pm0.1$. In addition, using deep data from the Australia Telescope Compact Array, we show that W41 is a faint ($9.8\pm1.4 \mu$Jy at 5.5 GHz) steep spectrum ($\alpha = -1.8\pm0.4$) radio continuum source, as expected for a pulsar. Light curve modelling of Hubble Space Telescope photometry shows evidence for ellipsoidal modulation with mild irradiation of a $\sim 0.5$--$0.55 M_{\odot}$ secondary around an invisible companion, also consistent with the redback interpretation. The precise position of W41, along with its well-measured 10.4-hr orbital period, should enable it to be matched to a newly-discovered radio pulsar in future data. Our result shows that close pulsar binaries are still hiding, misclassified among active binaries, even in well-studied clusters like 47 Tuc.
\end{abstract}


\maketitle

\section{Introduction}

The first millisecond pulsar was initially identified as an object of interest based on its steep radio spectrum and the fact that it scintillates, and then was later confirmed with timing measurements \citep{Backer1982}.  Searches for globular cluster pulsars were motivated by the combination of the excess in the number of X-ray binaries in globular clusters relative to field star populations \citep{clark, Katz} due to dynamical formation mechanisms \citep{FPR1975,Hills,Verbunt1987} and the understanding that millisecond pulsars are recycled in low mass X-ray binaries \citep{Smarr,Alpar}.  The initial discovery of a millisecond pulsar in a globular cluster was made in the cluster M28, first via imaging \citep{Mahoney} and then confirmed by timing \citep{Lyne}, and this led to a series of imaging searches of globular clusters for point sources, both steep spectrum and otherwise, as well as timing searches for pulsars \citep{Johnston}.

In recent years, this process has begun to repeat itself, as radio telescopes have upgraded their bandwidths to become more sensitive and new facilities have been built. Image-plane searches for pulsars have turned up a new pulsar in the Galactic Center region \citep{2017MNRAS.468.2526B}, as well as a large number of additional candidates  \citep{2024MNRAS.531.2191P}, including further out in the Bulge \citep{Frail2024}. New imaging searches in globular clusters have led to the discovery of a pulsar in the highly reddened cluster GLIMPSE C01 \citep{McCarver}, as well as new candidate pulsars in Terzan 5 \citep{Urquhart2020,Urquhart2026} { and a pulsar localised via variability in 47 Tuc \citep{Heywood2023} that had previously been found, but not localised, with timing data as a redback system that only rarely has uneclipsed emission \citep{Camilo2000,2016MNRAS.462.2918R}}.

At the present time, in many of the nearest clusters, existing pulsar timing data are already quite sensitive.  It is most likely, then, that some of the pulsars which have not yet been identified are in binaries in which interactions between the pulsar wind and the donor star can lead to substantial column densities of ionised gas within the binaries, which effectively eclipse the pulsar.  This supposition is borne out by the fact that
many of the recent discoveries in these clusters are black widow or redback pulsars \citep{2013IAUS..291..127R}---collectively known as ``spider'' binaries.
An alternative possibility is that some of the yet-undiscovered binary pulsars are in binaries with unusual orbital parameters, such that their accelerations are larger than currently searched computationally.  For example, the highly eccentric binary pulsar in NGC~6652 was only found because of an intensive search motivated by Fermi-band GeV emission \citep{deCesar}, and more systems like it could be identified by radio continuum imaging, as well.  

Discoveries via continuum imaging can also lead directly to information about the systems' optical counterparts.  In the globular cluster NGC~6397, a steep spectrum radio source \citep{Zhao2020} was identified as a pulsar candidate.  It  was then found to be associated with a binary star with a measurable orbital period of just under 2 days \citep{Pichardo2021} and finally, making use of the knowledge of the orbit of the binary, radio pulsations were confirmed from the system \citep{Zhang2022}.  

A new generation of imaging surveys of globular clusters has been primarily motivated by the searches for intermediate mass and stellar mass black holes \citep{M04,M05,Strader2012a,millerJones2015,2018ApJ...862...16T}, and for this reason has been undertaken at higher frequencies (4--10 GHz) than normally used for pulsar searches ($\sim 0.5$--3 GHz). The most extensive of these surveys is MAVERIC
\citep{Shishkovsky2020,Tudor2022}, which observed 50 Galactic globular clusters with deep Karl G.\ Jansky Very Large Array (VLA) or Australia Telescope Compact Array (ATCA) imaging. Despite the non-ideal frequency coverage for pulsar searches, owing to the depth of the images, MAVERIC identified a large number of previously unknown steep spectrum radio sources in its clusters that await further optical and radio-timing follow-up. Even deeper data than MAVERIC have been obtained for 47 Tuc \citep{2024ApJ...961...54P} and $\omega$ Cen \citep{Mahida2026}, which are sensitive to faint pulsars in these clusters.

{ 47~Tuc has one of the largest and best-studied pulsar populations of any globular cluster.  It has 42 known pulsars, second only to Terzan~5 \citep{PauloGCpulsars}, most of which have been discovered in batches as new data sets and/or computational capabilities have been obtained \citep{Manchester90,Manchester91,Camilo2000,Lorimer2003,Pan2016,Ridolfi2021,Chen2026}. Extensive optical \citep{2001ApJ...557L..57E,2002ApJ...579..741E,Edmonds03,RS2015} and X-ray \citep{Bogdanov2006} follow-up has been done on the known pulsars, with several showing strong optical/UV variability that helped ensure that the right counterpart had been identified.  }

In this Letter, we present evidence that the object 47~Tuc~W41 \citep{Edmonds03}, located at an X-ray position of RA $= 00^{\rm{h}}24^{\rm{m}}04{\overset{\rm{s}}{.}}3351$ and Dec $= -72^\circ05'01\overset{''}{.}451$ \citep[\textit{Chandra} Source Catalog v2.1;][]{2024ApJS..274...22E}
which had 
previously been suggested 
to be a coronally active binary \citep{Albrow,Edmonds03}, is actually a pulsar binary.  We discuss implications for searches for additional systems of this nature.

\section{Data}

\subsection{Radio data}
The source is detected in deep radio imaging of 47 Tuc from ATCA. The data consist of about 480 hours of imaging of the cluster taken mostly in 2021 and 2022, with a noise level of about 1 $\mu$Jy/beam at both 5.5 and 9 GHz. The full data reduction and analysis procedure is outlined in \citet{2024ApJ...961...54P}.

\begin{figure*}
\centering
\includegraphics[width=6 cm]{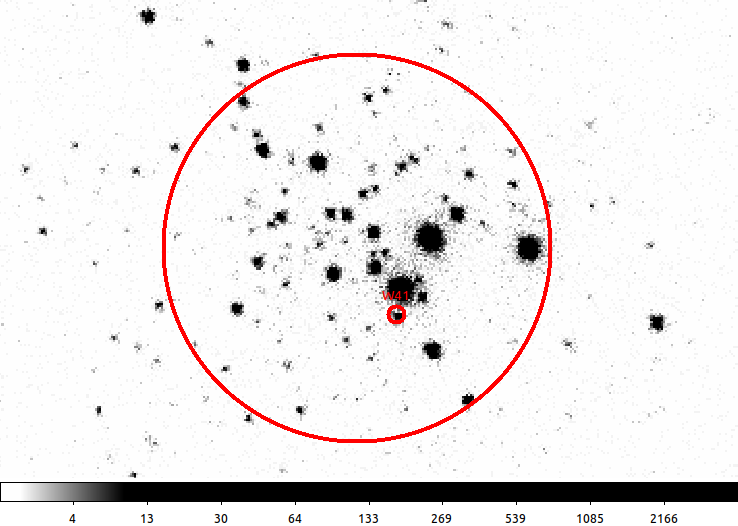}
\includegraphics[width=6cm]{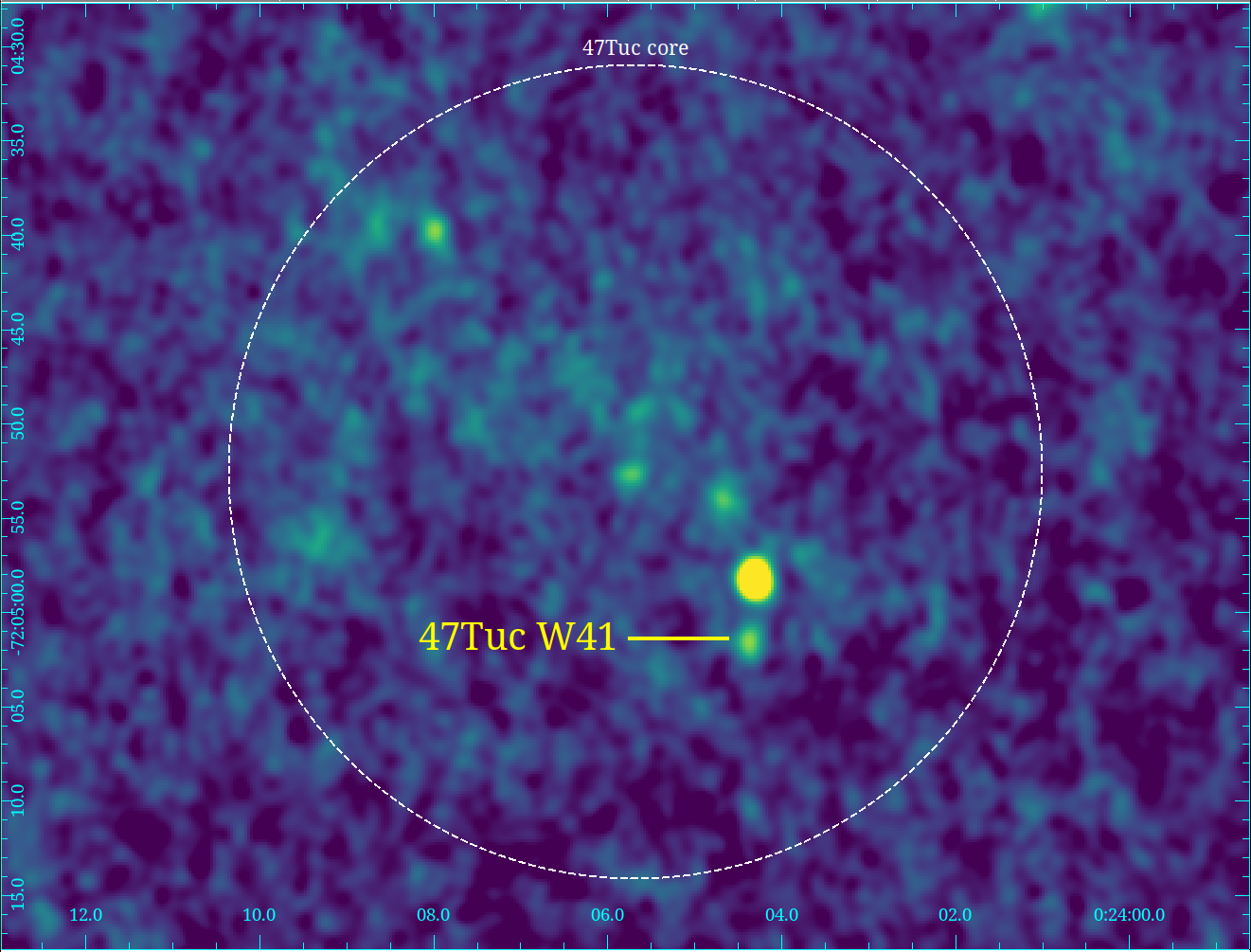}
\includegraphics[width=4.5cm]{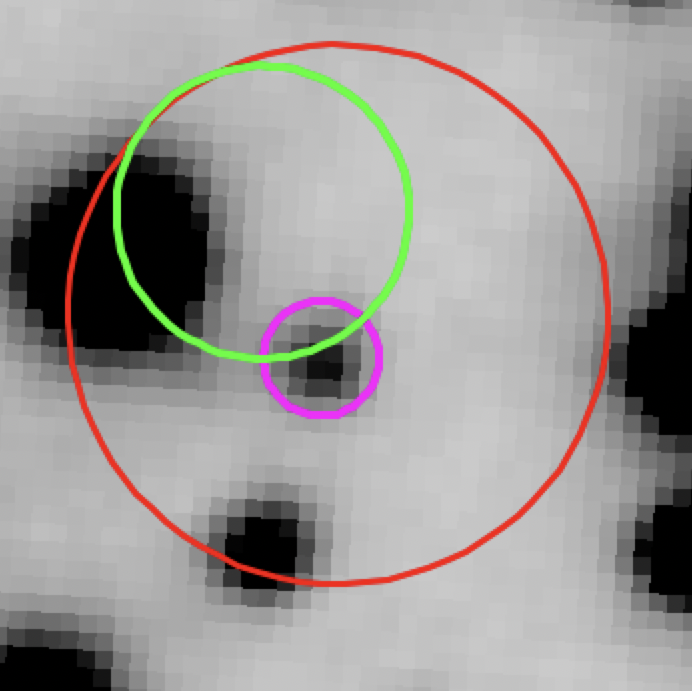}\\

\caption{Left: The summed X-ray image from Chandra, with the large red circle representing the core radius, and the smaller circle a 1" circle around W41.  Centre: the ATCA image of 47 Tuc showing the core radius and the location of W41. Right:Finding chart of W41 in the F300X filter, zoomed in considerably relative to the other two images. The red circle is the 95\% Chandra error circle as given in \cite{Bhattacharya2017}. The green circle represents the $1\sigma$ radio uncertainty . The magenta circle indicates the position of the F300X counterpart. }
\label{fig:chart}
\end{figure*}

The radio source is located at RA $= 00^{\rm{h}}24^{\rm{m}}04{\overset{\rm{s}}{.}}3471$ and Dec $= -72^\circ05'01\overset{''}{.}193$, with a localisation uncertainty of $0\overset{''}{.}07$ in the RA and $0\overset{''}{.}15$ in the Dec co-ordinate. This radio source is $0\overset{''}{.}26$ from the X-ray coordinates of W41, within the reported $0\overset{''}{.}3$ uncertainty circle of the X-ray localisation in the \textit{Chandra} Source Catalog (Figure \ref{fig:chart}). Given 75 X-ray sources within the 24" cluster core, the probability of an X-ray source lying within 0.26" of a radio source is only 0.9\%, so it is highly likely that the radio and X-ray sources are associated.

This source exhibits a likely steep spectrum, with flux densities of $9.8\pm1.4~\mu$Jy at 5.5~GHz and $3.9\pm0.9~\mu$Jy at 9~GHz. The best estimate of the spectral index for the source (where the spectrum scales as $\nu^\alpha$) is $\alpha=-1.8\pm0.4$, and a flat spectrum can be ruled out at $4\sigma$.  

\subsection{Optical and NUV data}
In this paper we have used data from the Hubble Space Telescope in the filters F435W (B), F658N ($H\alpha$) and F635W (R), which were acquired under programme GO 9281 with ACS-WFC in 2002 September 30, 2002 October 2/3 and 2002
October 11 (PI: Grindlay). We also used NUV data in the filters F390W and F300X acquired under the GO 12950 programme with WFC3 on August 13 2013 (PI: Heinke). Astrometry for the optical and NUV data was performed as in \cite{RS2015}. Photometry for the NUV images, together with extinction coefficients in the corresponding filters, was obtained as described in \cite{RS2018}. PSF photometry for the optical images was also obtained with the DOLPHOT software package \citep{Dolphin2016} using the individual FLC images.

\begin{figure}
\includegraphics[width=8.5cm]{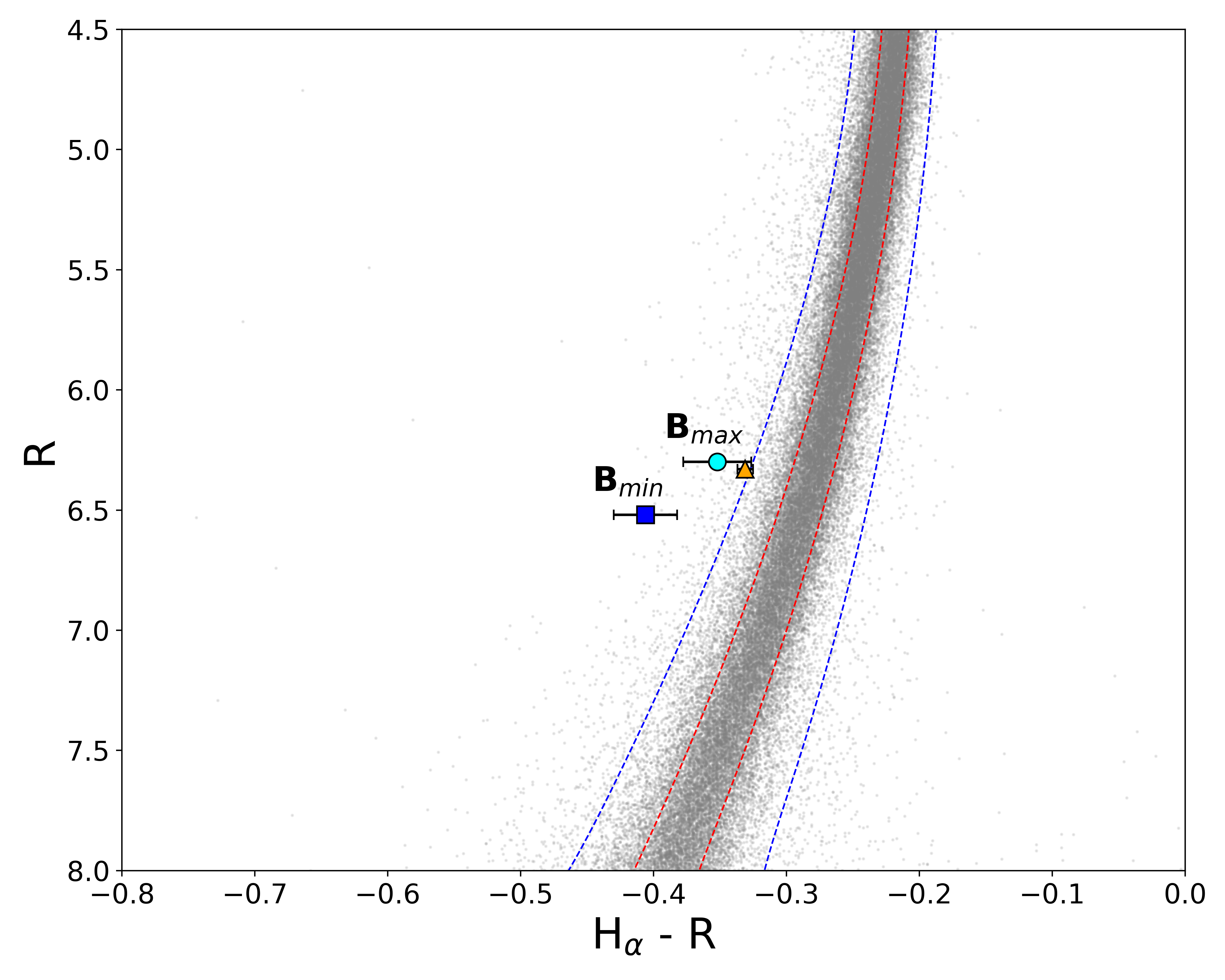}
\includegraphics[width=8.5cm]{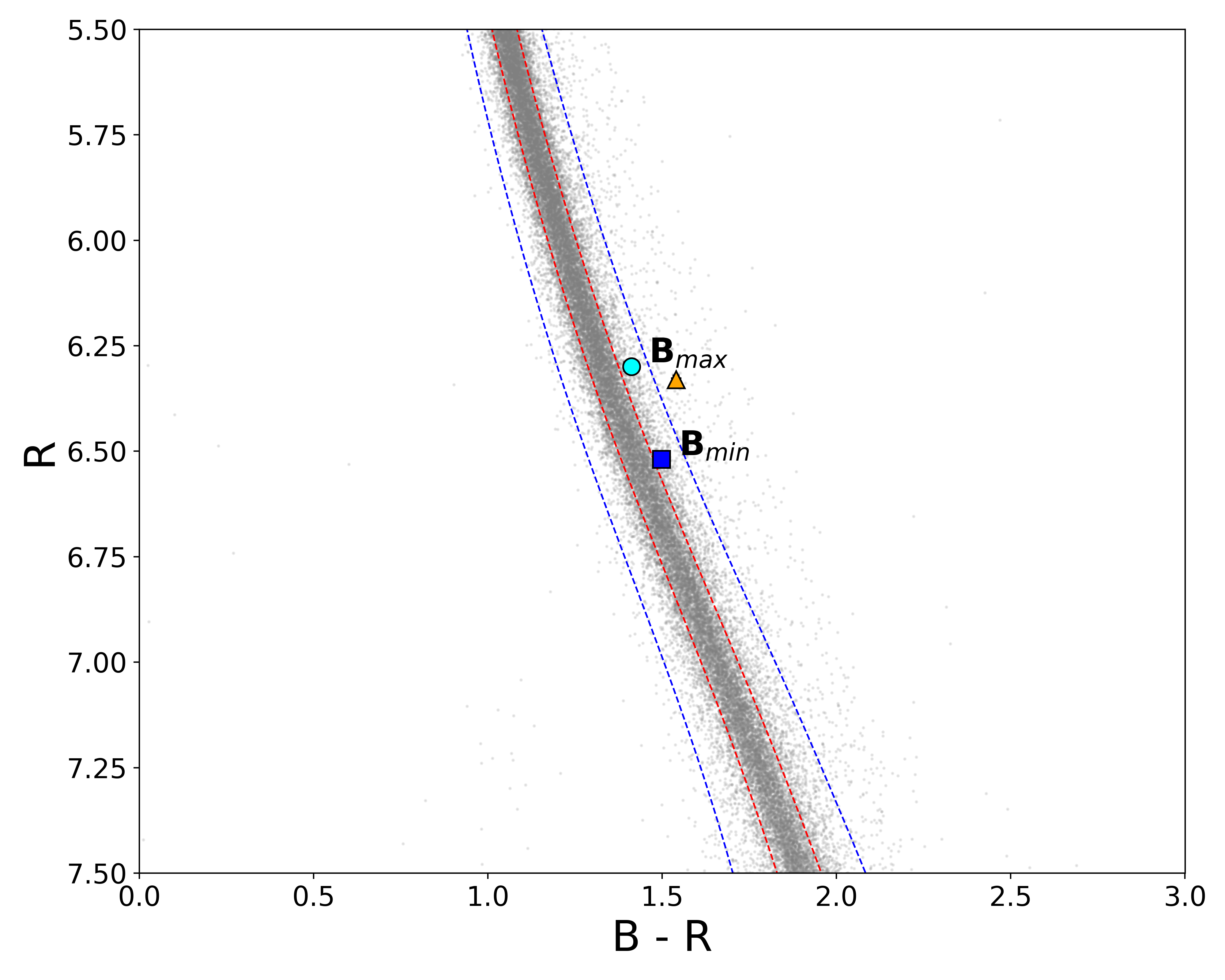}
\includegraphics[width=8.5cm]{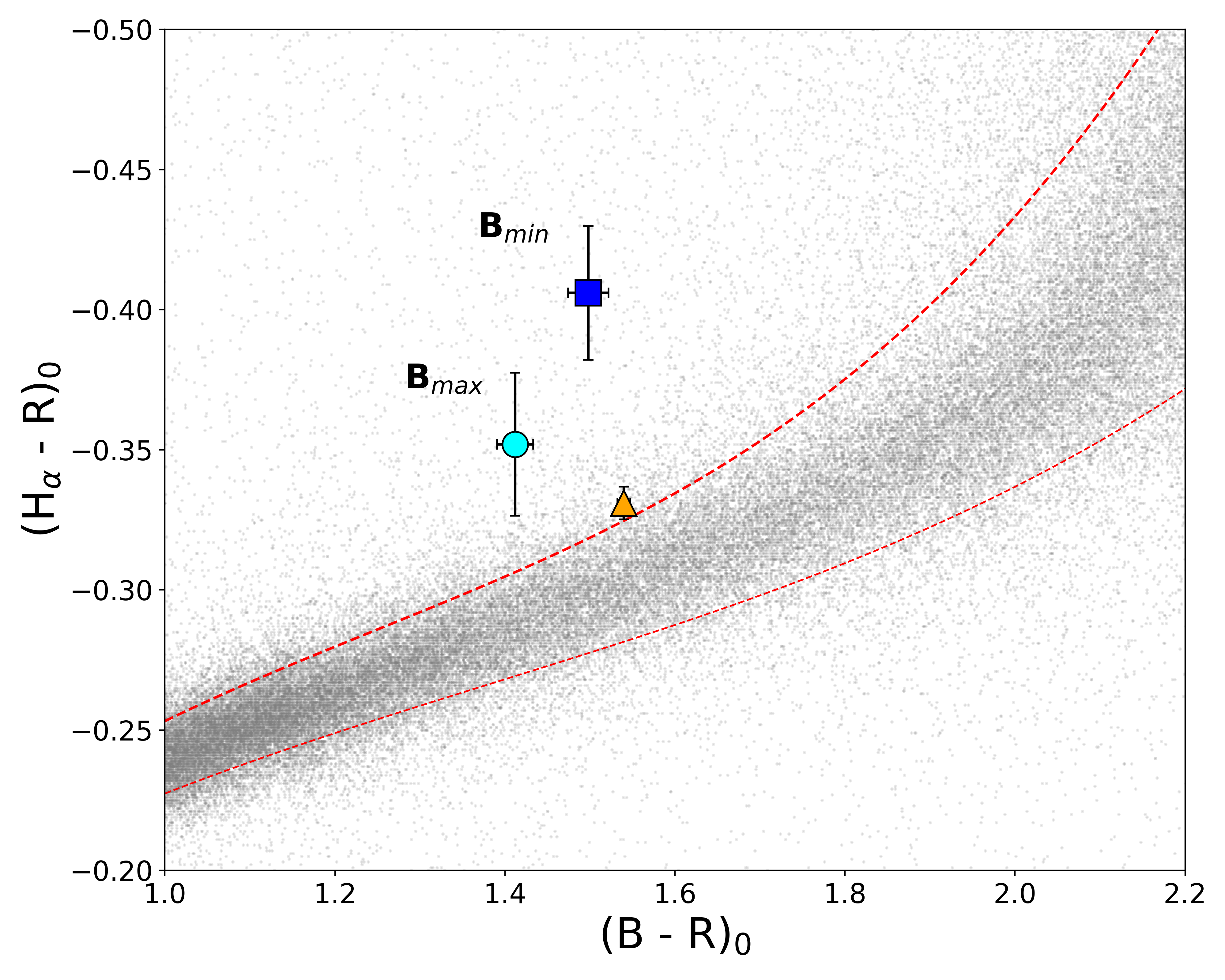}
\caption{Dereddened optical CMDs and color-color diagram of 47 Tuc. The blue and red lines around the main sequence indicate the 1 and 3 $\sigma$ dispersion limits from the main sequence ridgeline. The average magnitude of W41 is marked with an orange triangle. The blue square and cyan circle refer to points taken at the same phase, taking as a reference the min and max phase in the B band light curve. The system has an H$\alpha$ excess relative to other stars of the same B-R color, and a broader continuum, with an R band excess. 
}
\label{fig:CMDs}
\end{figure}

\begin{figure}
\includegraphics[width=8.5cm]{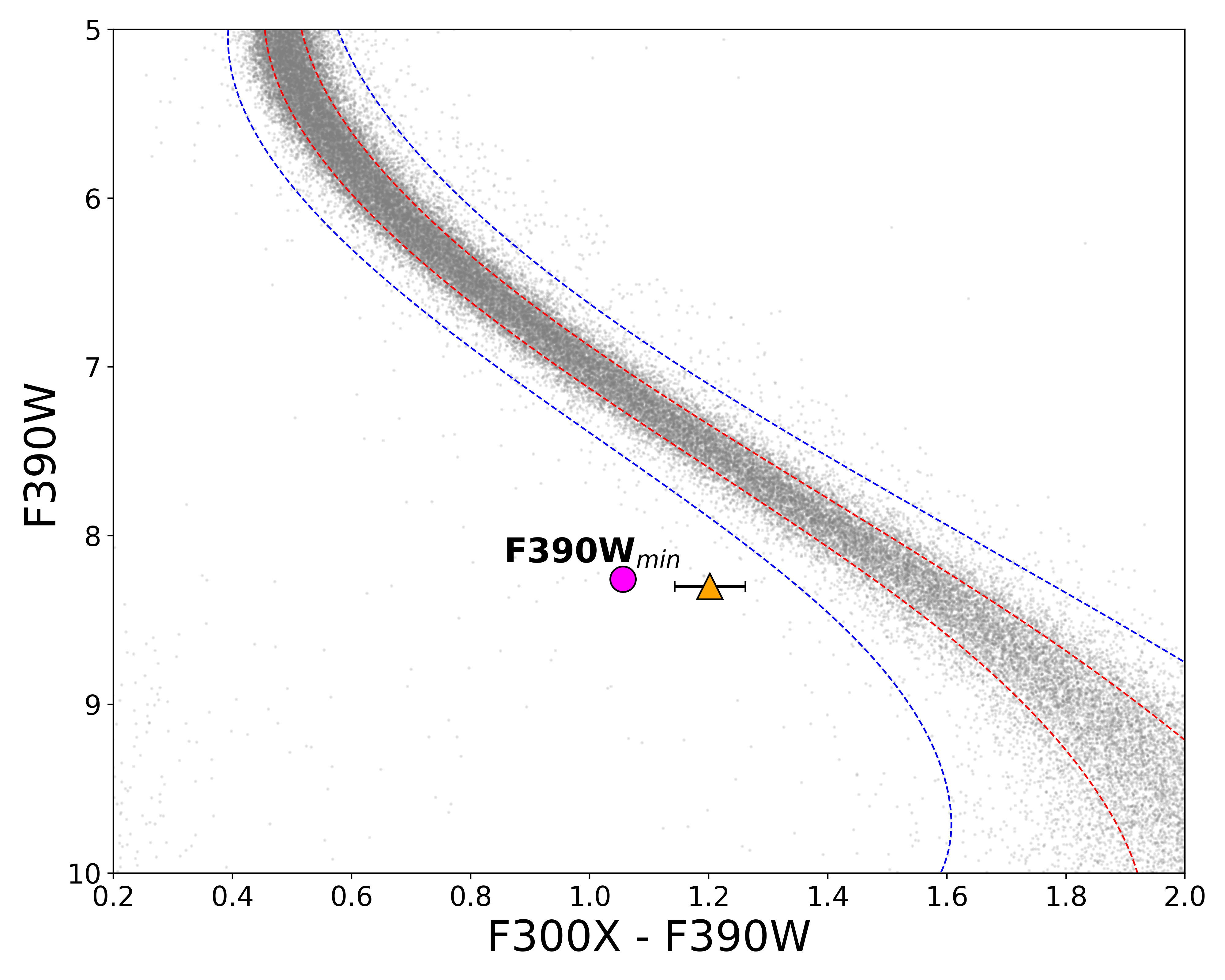}
\caption{Dereddened F300X - F390W vs F390W CMD of 47 Tuc. The average magnitude of W41 is marked with an orange triangle. The magenta circle refers to points taken at the same phase, taking as a reference the min phase in the F390W band light curve.}
\label{fig:NUV_CMD}
\end{figure}

Color magnitude diagrams (CMD) with the position of W41 in the different bands are shown in Figures \ref{fig:CMDs} and \ref{fig:NUV_CMD}. The binary shows a small H$\alpha$ excess in the H$\alpha$-R vs R CMD and on the color-color diagram. It was clearly detected on the binary sequence of the B-R vs R CMD. In the NUV, the binary shows a blue color which indicates a substantial contribution from a hotter component. 

\begin{figure}
\includegraphics[width=8.5cm]{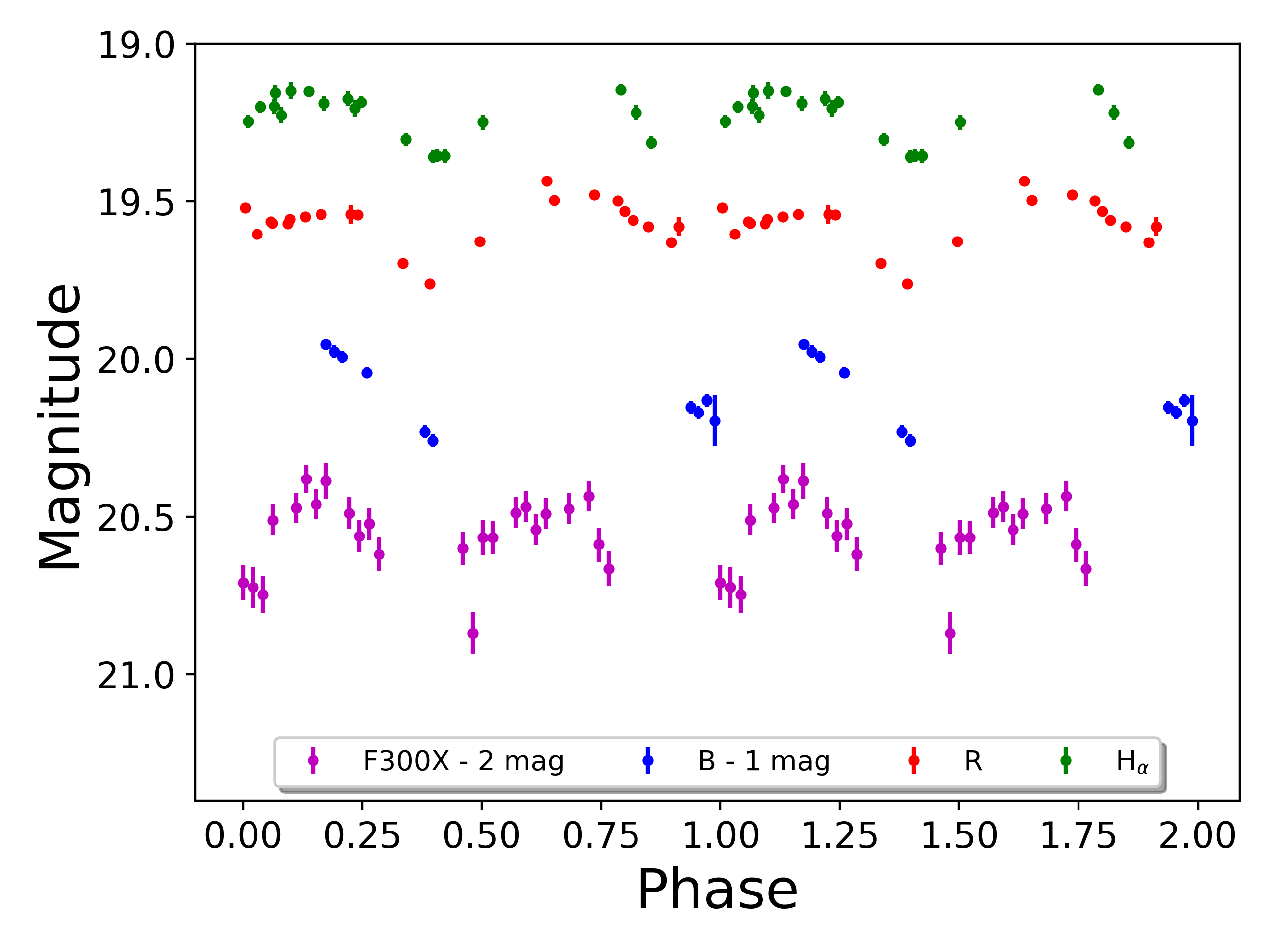}
\caption{Dereddened light curves of the optical counterpart to W41. These light curves were folded at the $0.4145$~d period identified by \citet{Albrow}. For clarity, the phase is plotted twice.}
\label{fig:LCfilters}
\end{figure}

Light curves in the different filters are presented in Figure \ref{fig:LCfilters}. They were folded at the 0.4145 d orbital period found by \cite{Albrow}, using as T0 the date of the first measurement in the F300X data. All measurements obtained by Dolphot are displayed. The modulation is observed in all filters.

\subsection{X-ray data}
47~Tuc has been extensively observed by \chandra\ (more than 500 ks with ACIS and 700 ks with HRC). While W41 is relatively faint, it is clearly detected in most \chandra\ observations. Thus, we extract spectra from all \acis\ observations with exposure $\geq 10$~ks (to ensure reliable spectral extraction). 

All \chandra\ data were reduced and reprocessed using \textsc{ciao} \citep[version 4.14 CalDB 4.9.8,][]{Fruscione2006}, with spectra extracted via \texttt{specextract} assuming  a circular extraction region with a radius of $1.1''$ centred on the source, and a background annulus with inner and outer radii of $2''$ and $10''$, respectively, with all detected sources within and close to this region excluded from extraction. The background selection is to make sure to consider the elevated background levels in the core of 47 Tuc due to crowding. We then combined the extracted spectra (and associated files) with \texttt{combine\_spectra}, and binned spectra with at least one background count per bin.

\hrc\ data were first barycenter-corrected (using \texttt{axbary}), re-projected and stacked (with \texttt{reproject\_obs}). Background-subtracted light curves were extracted using \texttt{dmextract}, assuming Gehrels uncertainties \citep{Gehrels1986}.  The light curve from the \chandra\ data is shown folded on the orbital period in Figure \ref{fig:xraylc}.  It shows statistically marginal evidence for this periodicity.

\section{Analysis}

\subsection{X-ray spectral analysis}
\begin{figure*}
\includegraphics[width=8cm]{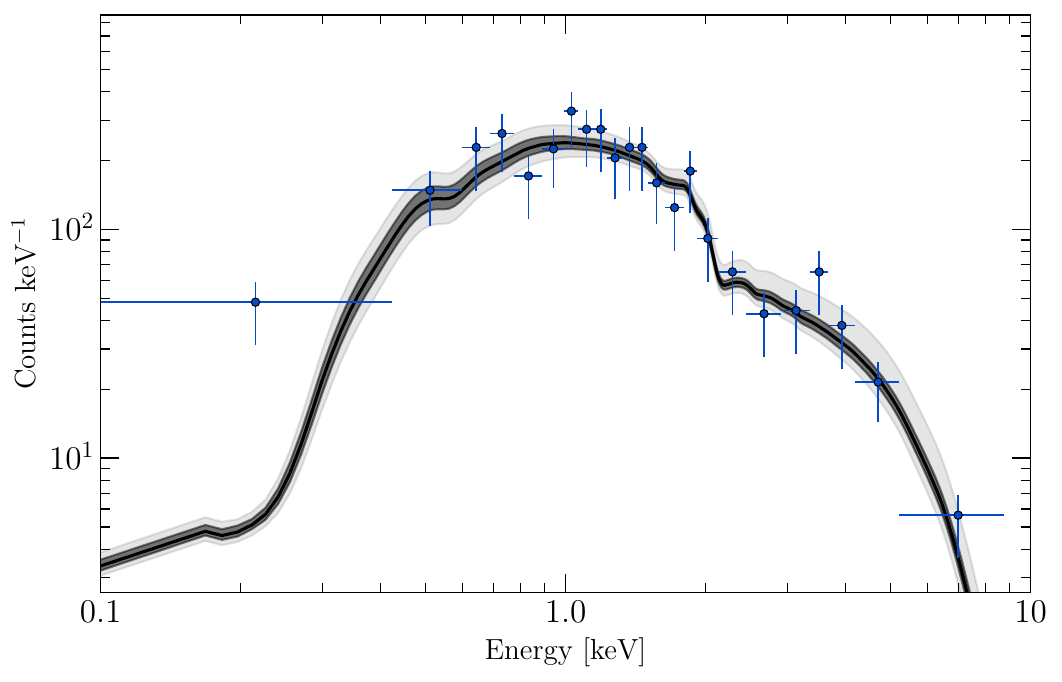}
\includegraphics[width=8.5cm]{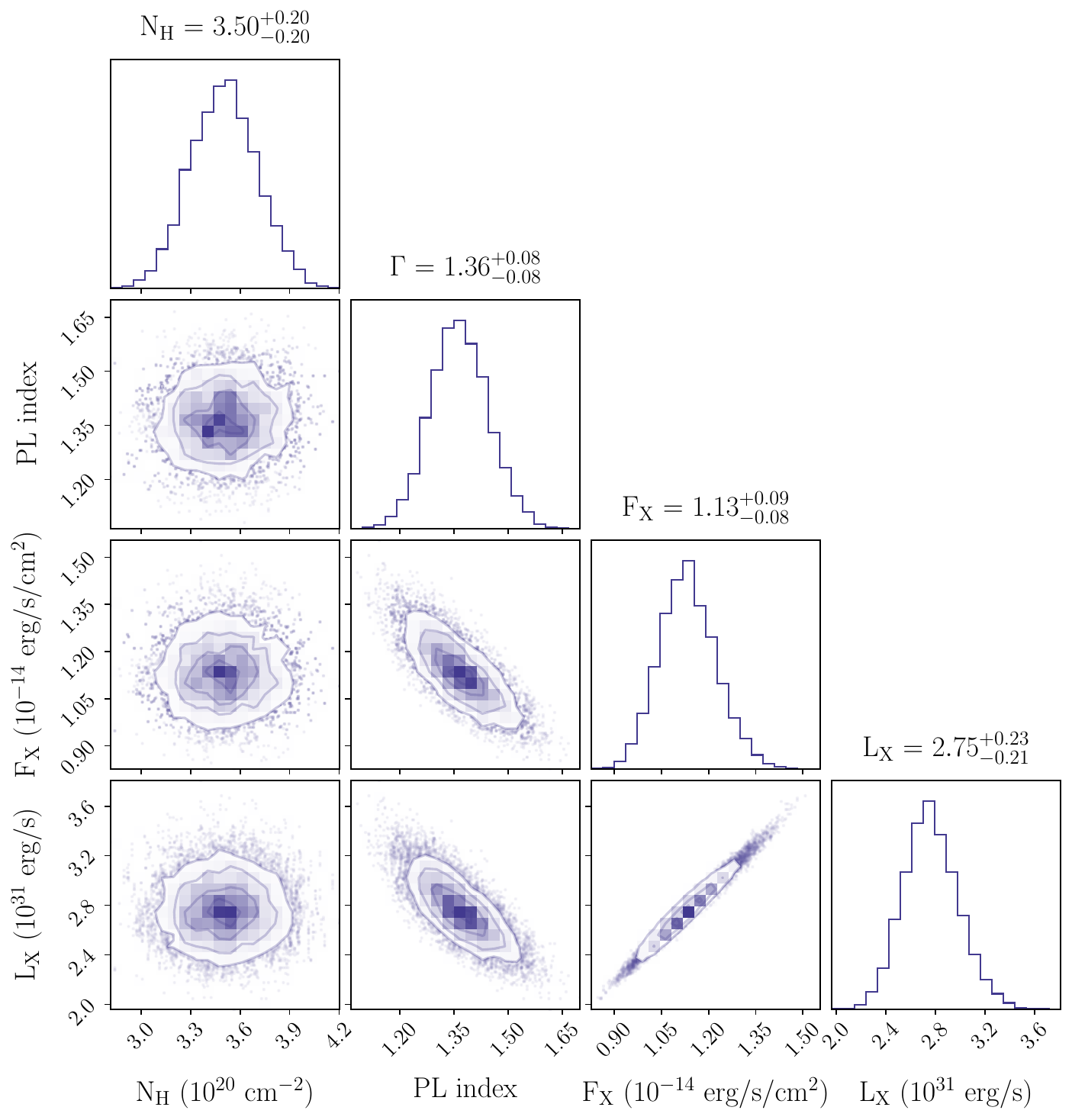}
\caption{{\it Left}: \acis\ X-ray spectrum of W41 (blue circles), model posterior median (solid black line) and 90\% highest density interval (shaded gray region). Here, the spectrum is grouped by 20 counts per bin for plotting. {\it Right}: Pairwise plot of posterior samples. Unabsorbed flux (F$_{\textrm{X}}$) is reported in the 0.5$-$10 keV band. X-ray luminosity (in the same band) is inferred from the flux assuming a distance of $4.52\pm0.03$~kpc to 47Tuc \citep{Baumgardt2021}.}
\label{fig:xrayspec}
\end{figure*}

\begin{figure}
\includegraphics[width=8.5cm]{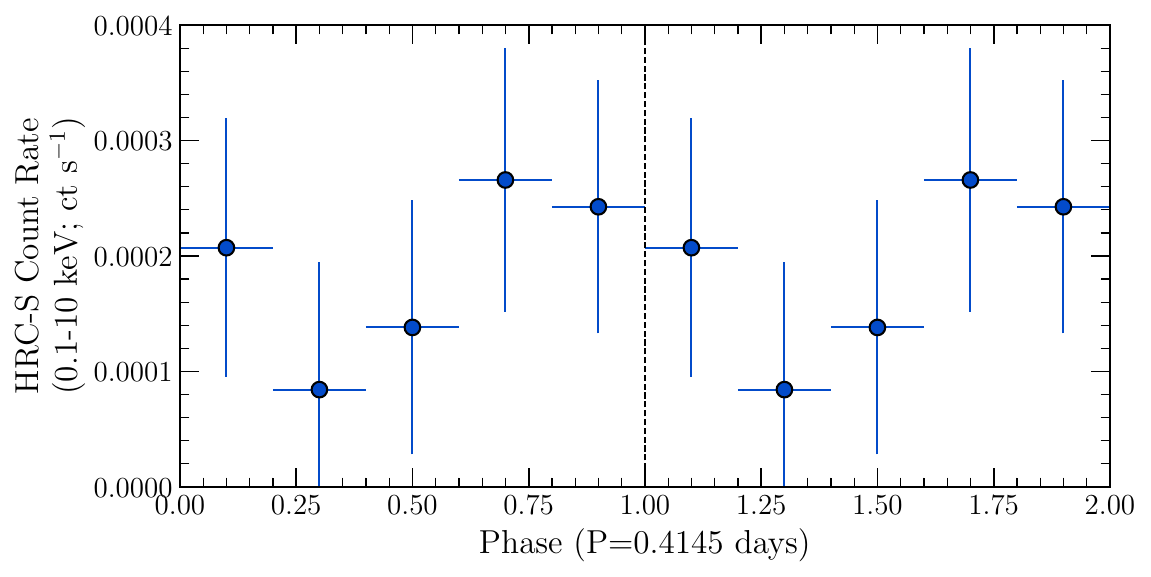}
\caption{\hrc\ light curve of W41, folded on the period reported by \citet{Albrow}. While there is weak evidence for a similar periodic modulation in the X-rays, it is not statistically significant. 
}
\label{fig:xraylc}
\end{figure}

We performed spectral analysis with \textsc{Xspec} \citep[version 12.12.1c,][]{Arnaud1996} and \textsc{Bxa} \citep[version 4.0.5,][]{Buchner2014} with the $w$-statistic \citep[][assuming a Poisson process for the background]{Cash1979}, abundances of elements based on \citet{Wilms2000}, and photo-electric cross-sections based on \citet{Verner1996}. Modelling the spectrum with an absorbed power-law model (\texttt{tbabs*pegpwrlw} in \textsc{Xspec}), we assumed the following priors: a Gaussian prior on hydrogen column density (N$_{\textrm{H}}$) based on the hydrogen column density towards 47~Tuc \citep[$3.5\pm0.2\times10^{20}$ cm$^{-2}$,][]{Salaris2007, Bahramian2015, Foight2016}, uniform prior on photon index ($\Gamma$) with lower and upper bounds of 0 and 5, respectively, and a log-uniform prior on power-law normalization (i.e., unabsorbed flux in the 0.5-10 keV band) with lower and upper bounds of $10^{-17}$ and $10^{-10}$ erg~s$^{-1}$~cm$^{-2}$. Posterior estimates for model parameters are tabulated in Table~\ref{tab:specparams}. The spectrum and pairwise distributions of the posterior samples are visualised in Figure~\ref{fig:xrayspec}.

\begin{table}
\centering
\caption{Posterior estimates (median, along with 16 and 84 percentiles) for the absorbed power-law model. Unabsorbed flux (F$_{\textrm{X}}$) is reported in the 0.5$-$10 keV band. X-ray luminosity (in the same band) is inferred from the flux assuming a distance of $4.52\pm0.03$~kpc to 47Tuc \citep{Baumgardt2021}. It is worth noting that the posterior sample for N$_{\textrm{H}}$ appears similar to the prior distribution, indicating the weak  influence of data on constraining N$_{\textrm{H}}$.}
\label{tab:specparams}
\begin{tabular}{lc}
\hline
\hline
Parameter                                       & Posterior estimate \\
\hline
N$_{\textrm{H}}$ (10$^{20}$ cm$^{-2}$)          & $3.5\pm0.2$        \\
$\Gamma$                                        & $1.36\pm0.08$      \\
F$_{\textrm{X}}$ (10$^{-14}$ erg~s$^{-1}$~cm$^{-2}$) & $1.13\pm0.09$      \\
L$_{\textrm{X}}$ (10$^{31}$ erg~s$^{-1}$)       & $2.7\pm0.2$        \\
\hline
\end{tabular}
\end{table}

\subsection{Light curve modeling and companion properties}

While the light curves plotted in Figure \ref{fig:LCfilters} are too sparse for good constraints for light curve fitting, there is an additional data source: the phased light curve in $V$ and $I$ published in Figure 12 of \citet{Albrow}. While this light curve has the advantage of a large number of data points (over 600 in each filter), it also has a disadvantage: these data are not available in tabular form, nor do they have the original epochs attached---they are only available in the \citet{Albrow} phasing.

Given these constraints, we decided to model this light curve in a preliminary manner. We digitised the data points, applied a correction for foreground extinction, and then began by fitting a pure ellipsoidal model using {\tt PHOEBE} version 2.4.22 \citep{Prsa2005,Conroy2020}. Since light curve fitting alone cannot constrain the mass of the primary independent of the secondary, we fixed the primary mass to $1.8 M_{\odot}$, a typical value for redbacks \citep{Strader2019}, [Fe/H] = --0.5 (reasonable for a solar-scaled abundance given that 47 Tuc has [Fe/H] = --0.7 and is $\alpha$-enhanced; \citealt{Alves-Brito2005}), and a distance of 4.52 kpc \citep{Baumgardt2021}. We fit the secondary mass and $T_{\rm eff}$, the Roche lobe filling factor, the binary inclination, and a phase offset (no specific conjunction appears to have been chosen for the phase zeropoint for the figures in \citealt{Albrow}). The normalisation of the models is \emph{not} a free parameter---it is set entirely by the distance assumed and the shape and temperature of the secondary.

\begin{figure}[t]
\includegraphics[width=8.5cm]{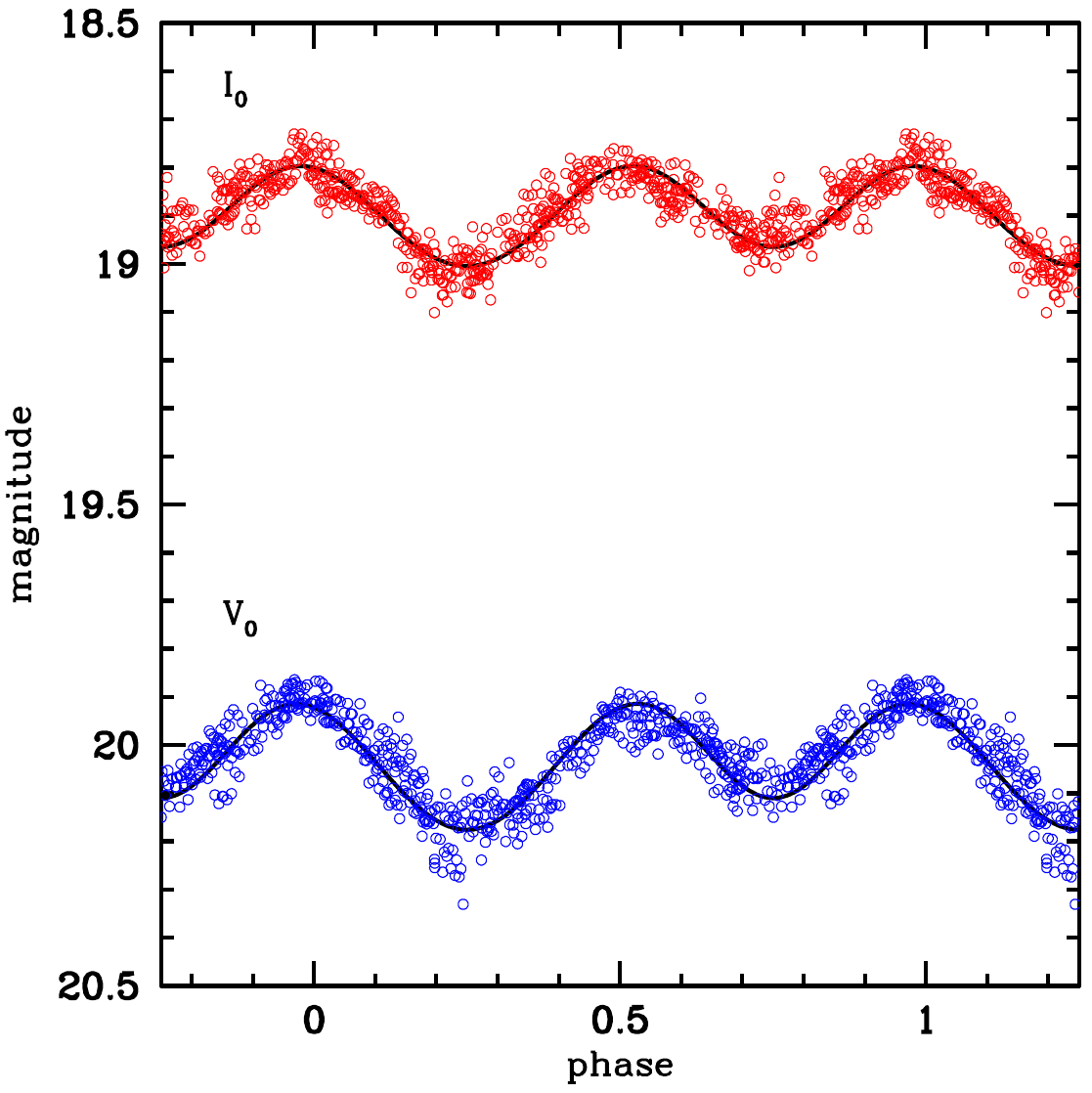}
\caption{Irradiated ellipsoidal light curve fit to HST/WFPC2 $V_0$ and $I_0$ photometry of W41, as described in the text. We adopt the pulsar phase convention where $\phi=0$ is the ascending node of the pulsar, so $\phi=0.25$ is the inferior conjunction of the secondary.}
\label{fig:lc_mod}
\end{figure}

The best-fitting ellipsoidal model has $M_2 = 0.50 M_{\odot}$, $T_{\rm eff} = 4500$ K, a Roche lobe filling factor of 0.96, and a binary inclination of $62^{\circ}$. The overall quality of the model fit is not great ($\chi^2/d.o.f .\sim 1.7$), and the the residuals of this fit are dominated by a single broad single peak, which is characteristic of irradiation.

We then add a central irradiating source, with its luminosity as another free parameter. While this luminosity is not amenable to direct interpretation, it can serve as a stand in for the luminosity of the pulsar wind, some of which is expected to heat the companion; irradiation from an intrabinary shock could also be relevant if the shock is close to the secondary. 

In this interpretation of the light curve; the phase is shifted by $\phi = 0.5$ compared to the pure ellipsoidal model, such that the deepest minimum is when the nightside of the secondary is visible, while the dayside (which would normally be the faintest due to gravity darkening) is instead slightly brighter due to the irradiation. The best-fit stellar parameters in this model are very similar to the pure ellipsoidal model ($M_2 = 0.55 M_{\odot}$, $T_{\rm eff} = 4500$ K, filling factor $0.93$), but with a more edge-on inclination required ($i = 85^{\circ})$. The implied volume-equivalent radius of the star is $0.82 R_{\odot}$.  Accommodating such a high inclination angle, along with the lack of orbital modulation of the X-ray emission, would require a rather large intrabinary shock region;  the pulsar 47~Tuc~W is thought to have a high binary inclination and shows only sporadic modulation of its X-rays on the orbital period \citep{2021MNRAS.500.1139H}, so such a scenario is not an unreasonable one.

This new fit is plotted in Figure \ref{fig:lc_mod}. It is a much better fit overall ($\chi^2/d.o.f. \sim 1.2$), representing an excellent fit to the $I_0$ light curve and an adequate fit to the $V_0$ light curve. Given how common irradiation is in redback light curves, it seems reasonable to expect that it is present here as well. 

Under the assumption that the secondary is not overfilling its Roche lobe (reasonable given the evidence for a radio pulsar, which would not be visible if accretion were present), we can calculate a minimum mass of the secondary, since we have a measurement of its radius, and the mean density of a Roche lobe-filling star is determined almost entirely by the orbital period and is insensitive to the primary mass \citep{Paczynski}.
Our radius measurement of $0.82 R_{\odot}$ implies a minimum mass of $\sim 0.45 M_{\odot}$. We can also roughly estimate the maximum mass of the secondary by comparing the absolute magnitude to that of normal main sequence stars in 47 Tuc, under the assumption that an irradiated secondary should not be less luminous than a main sequence star of the same mass. The mean nightside minimum $V_0 = 6.88$ corresponds to a main sequence mass of about 0.62 $M_{\odot}$, using a MIST isochrone \citep{Dotter2026} with [Fe/H] = --0.7, [$\alpha$/Fe = +0.2], and age 12.5 Gyr, appropriate for 47 Tuc. These masses encompass the values found by the light curve fitting.

Given the issues with the provenance of the data themselves, we do not argue that the exact derived values, especially for the inclination and Roche lobe filling factor, should be take too seriously. Nonetheless, the properties are typical for redback systems. The fit is a plausibility check that shows the interpretation suggested by the rest of the data are are also broadly consistent with these unusually well-sampled light curves.

\section{Discussion}
The object W41 in 47~Tucanae has previously been classified on several occasions as a coronally active binary, either as a member of the BY~Draconis class, or a semi-detached W~Ursa~Majoris  \citep{Albrow}.  It was noted that this object has an X-ray luminosity higher than expected for these classes of objects \citep{Edmonds03}, and that, unusually for active binaries, it showed no statistically significant variability \citep{Heinke2005}.  Steady coronal X-ray activity saturates at 0.1\% of the bolometric luminosity \citep{VilhuWalterSaturation}.   \citet{Edmonds03} suggested that it might have been caught in a flaring state by Chandra, since flaring can bring the X-ray emission above the saturation level, and the source otherwise showed consistency with semidetached 
W Ursa Majoris variables \citep{Albrow}. As the subsequent Chandra data 
continue to show a similar $L_X$, it is clear that flaring cannot explain  the source exceeding the saturation limit. Given the X-ray excess, the source would be most likely either to have accretion or to be a pulsar binary.  The steep spectrum radio emission strongly favours the latter possibility. 

\subsection{W41 as a redback pulsar}

All data for W41 are consistent with a single interpretation: that it is a redback millisecond pulsar. The faint, likely steep-spectrum radio emission is as observed for pulsars. The 0.5--10 keV X-ray luminosity ($3 \times 10^{31}$ erg s$^{-1}$) and hard photon index ($\Gamma = 1.4\pm0.1$) are fully consistent with that observed for redbacks \citep{Roberts2015,Urquhart2020}, where the X-rays arise primarily in an intrabinary shock. The orbital period (10.4 hr) and inferred secondary mass ($\sim 0.5$--0.55 $M_{\odot}$) are likewise typical for redbacks \citep{Strader2019,Koljonen2025}. The light curve is well-fit by an ellipsoidal model with some irradiation, which is again normal for redbacks (e.g., \citealt{Turchetta2023}). The H$\alpha$ excess and possible UV excess observed in the HST photometry can be attributed to either ionization of the outflow of the secondary or the intrabinary shock, as observed in other redbacks (e.g., \citealt{Sabbi2003,Halpern2017,Strader2019}).

\subsection{Future characterization of the source}

Ultimate confirmation that W41 is a pulsar would come from its detection as a radio pulsar. New pulsar discoveries in 47 Tuc are ongoing as part of the Transients And PUlsars with MeerKAT (TRAPUM) project, with 15 new discoveries just reported \citep{Chen2026}. Of the known pulsars with measured binary periods or locations, none match W41. However, it is possible that it could be one of the new poorly localised binary pulsars: ah, an, aq, ar, or as, or else a yet-undiscovered pulsar. As timing observations of 47 Tuc are ongoing, future observations could confirm such an association.

Progress could also be made in the optical. While existing archival MUSE data are not sensitive enough to yield spectra of this star (S. Kamann, priv. comm.), with narrow-field mode adaptive optics observations, it should be straightforward to measure its radial velocity curve with MUSE or with the James Webb Space Telescope's integral field unit.

\subsection{Implications for other potential source misclassifications}
It is important to note that in the original classification of W41 as an active binary, \citet{Edmonds03} had already recognised that this system had an X-ray luminosity significantly higher than expected for an active binary and noted that the most likely explanation for this was flaring activity in the X-ray data.   Our attention was drawn to this source only because of the steep spectrum radio source.  It is likely then, that in clusters without such deep radio data, there are other spider pulsars that have been misclassified as active binaries for the same reason and which are waiting to be discovered.

\section{Acknowledgments}
The Australia Telescope Compact Array is part of the Australia Telescope National Facility (grid.421683.a) which is funded by the Australian Government for operation as a National Facility managed by CSIRO. We acknowledge the Gomeroi people as the traditional owners of the Observatory site. We thank S. Kamann for carefully investigating the VLT/MUSE IFU data for any sign of W41. 
AP thanks Texas Tech University for its hospitality during a visit in which this project began. C. O. Heinke is supported by NSERC Discovery Grant RGPIN-2023-04264. J.S. is supported by NSF grant AST-2205550 and NASA grants 80NSSC21K0628 and 80NSSC26K0226. L.C. is grateful for support from NASA grant 80NSSC23K0497. L.R.S. is grateful for support from grant CS-CSA2025-051 from the Cottrell Scholar program of the Research
Corporation for Science Advancement.

\section{Data availability}
The Australia Telescope Compact Array data are all public data which have been archived by the Australia Telescope National Facility.  The X-ray data presented are all public data, which have been archived by the Chandra X-ray Observatory.  The Hubble Space Telescope data are all public data which have been archived at the Space Telescope Science Institute.



\bibliographystyle{plainnat}
\bibliography{newbib}

\end{document}